# Experiments with Different Indexing Techniques for Text Retrieval tasks on Gujarati Language using Bag-of-Words Approach

Dr. Jyoti Pareek, Hardik Joshi, Krunal Chauhan, Rushikesh Patel

*Abstract*—This paper presents results of various experiments carried out to improve text retrieval of gujarati text documents. Text retrieval involves searching and ranking of text documents for a given set of query terms. We have tested various retrieval models that uses bag-of-words approach. Bag-of-words approach is a traditional approach that is being used till date where the text document is represented as collection of words. Measures like frequency count, inverse document frequency etc. are used to signify and rank relevant documents for user queries. Different ranking models have been used to quantify ranking performance using the metric of mean average precision. Gujarati is a morphologically rich language, we have compared techniques like stop word removal, stemming and frequent case generation against baseline to measure the improvements in information retrieval tasks. Most of the techniques are language dependent and requires development of language specific tools. We used plain unprocessed word index as the baseline, we have seen significant improvements in comparison of MAP values after applying different indexing techniques when compared to the baseline.

## I. Introduction

INFORMATION RETRIEVAL (IR) is defined as the technique of retrieving data/documents that are relevant to an information need. It is usually concerned with searching, manipulating and representing large collection of electronic text data. Information retrieval tasks have been executed since nearly six decades[1], with the evolution of data representation techniques, numerous techniques have developed to fulfil information needs. IR systems process information needs of any user, it does not restrict itself to text data, it may include image, audio or video formats[2]. Text retrieval is the discipline that deals with retrieval of unstructured or partially structured text data, especially textual documents. The relevant documents are retrieved in response to a set of query which itself may be structured or unstructured. The typical interaction between a user and an IR system can be modeled as the user submitting information needs in the form of queries to the system; the system returning a ranked list of relevant documents that matches the queries. The ranked list of documents is ordered such that the most relevant documents are at the top of the list. When the information need is not known in advance and in the situation where the user query is fired once on an indexed data, the task is referred as ad hoc information retrieval [3].

The need for effective methods of automated indexing and automated IR has grown due to tremendous explosion in the amount of text documents and increase in the sources of information over the Internet. In the last decade, there has been a significant growth in the amount of text documents in Indian languages. Researchers have been performing IR tasks in English and European languages since many years through evaluation forums like TREC[4], CLEF[5] etc., efforts are being made to encourage IR tasks for the Indian languages through evaluation forum FIRE[6].

IR evaluation forums and research communities uses resources know as test collection[7]. The classic components of a test collection are:

1) A huge corpus that includes collection of text documents; a tag "docid" is used to identify each document uniquely.

2) A set of queries (also referred as topics); a tag "qid" is used to identify each query uniquely. The query is further classified as T (title), TD (title and description) and TDN (title, description and narration).

3) A collection of query relevance judgements (also referred as qrels or relevance judgement) which consists of pairs detailing the matching documents for each query, that is gold standard for each query.

Ad hoc IR can be represented as deriving a ranked list of the most relevant documents among a static collection of documents with regards to one time information need in the form of a query. A scoring function (a.k.a. retrieval model) is used to estimate the relevance and rank of each document among the document collection with reference to the query. In

Manuscript received December 14, 2016. This work was supported by the University Grants Commission (UGC) Minor Research Project, grant number: F. No: 41-1360/2012 (SR).

Dr. Jyoti Pareek is a Professor with Department of Computer Science, Gujarat University, India.

Hardik Joshi is an Asst. Professor with Department of Computer Science, Gujarat University, India (e-mail: hardikjoshi@gujaratuniversity.ac.in).

Krunal Chauhan is a Sr. Software Engineer at LogiCeil Solutions, Ahmedabad, India and Rushikesh Patel is a Solution Analyst at Canada Technology Partners Ltd., India.

bag-of-words approach, the document is taken into account as a collection of words, the semantic information like co-occurrence of words or linguistic information like parts of speech etc. are not taken into account.

## II. RETRIEVAL MODELS

In information retrieval, the query itself is represented as a document that may share the same document representation as the documents within the collection. Therefore, the relevance of any document can be interpreted as a measure of similarity between two documents (document belonging to the document collection and query document). In the case of bag-of-words approach, the document relevance is aggregated from the relevance of each query term taken separately. It is usually defined as the sum of each query term's weights in the query document and the collection. The IR task is to judge that how much each query term contributes to the overall relevance of the document belonging to the document collection. So, the documents matching more query terms that the others should be favored. However, large documents may contain more query terms, to resolve such cases; various statistical measures have been proposed to penalize large documents to some extent as they have more chance of containing a query term.

Several retrieval models have been proposed to improve on indexing and retrieval tasks. In this paper, we have experimented with Guajarati document collection using Terrier tool[8] that has implemented few widely used retrieval models like vector space model (TF-IDF model)[9], probabilistic models like BM25[10], language models like drichlet prior[11], information based approaches[12] and the divergence from randomness framework[13]. These are the state-of-art methods to perform ad hoc IR using bag-of-words approach.

## III. EXPERIMENTAL SETUP

Gujarati language is resource constrained language. To the best of our knowledge, there is a single corpus available to perform ad hoc IR tasks. An IR task must use Gold Standard data to evaluate various tools and techniques. Details of document collection (corpus) and topics (queries) are as following:

### A. Document Collection (Corpus)

To conduct our experiments, we used the collection of Gujarati text documents that were made available by Forum for Information Retrieval and Evaluation (FIRE) in 2011[6]. Details of Gujarati test collection used for experiments are mentioned in Table I. The test collection was created from the news article of the daily newspaper, "Gujarat Samachar", where the articles are included from 2001 to 2010. Each news article represents a unique document in our test collection. Statistical summary of the Gujarati document collection is given in Table

1. The test collection is available in Unicode text format (UTF) where each article is marked up using the following tags:
<doc> and </doc> : Tags to represent beginning of the document and end of document.
<docno> </docno> : Tags to uniquely identify the document.
  : Tags to represent the content of document.

TABLE I
STATISTICS OF GUJARATI TEST COLLECTION (CORPUS)

| Particulars | Quantity |
| --- | --- |
| Approx. Size of Collection | 2.7 Giga Bytes |
| Approx. Number of text Documents | 3,13,000 |
| Number of unique Terms | 20,92,000 (Approx..) |
| Number of Tokens | 13,92,73,000 (Approx.) |
| Average Number of Tokens per Document | 445 |

### B. Queries (Topics)

IR models were tested against 50 different queries in Gujarati language available as topic collections for FIRE exercise[6]. On the lines of TREC evaluation exercise[14], each query made available through FIRE is divided into three sections: the title (T) which indicates title of the query, the description (D) that gives a one-two sentence description of query and the narrative part (N), which specifies the relevance assessment criteria for a particular query. The following is an example of a single query document in the collection of queries:
<top>
<num> </num> : Unique identifier of the Query
<title> </title> : Title of the Query
<desc> </desc> : Short description of the Query
<narr> </ narr> : Detailed Query in narrative form
</top>

Each query document is represented using UTF format, the ad hoc retrieval tasks in FIRE[15] released a set of 25 queries for the year 2011 and another 25 queries for the year 2012. Different experiments were carried out on a total of 50 queries made available and subsequently the results were evaluated against the qrels (query relevance judgment) released under these evaluation exercises.

### IV. EXPERIMENTS WITH INDEXING TECHNIQUES

Although IR tasks are language independent, it is observed that they do not suite well with the Indian Languages. As most of the Indian languages are morphologically rich, when information retrieval is performed using language specific notion, better results are obtained[16]. We performed four different indexing experiments on Gujarati test collection. Most of the experiments were conducted using open source text retrieval tool, Terrier[8]. However, Gujarati being a resource constrained language, few resources like stop words list, stemming rules and fcg rules were created with the help of domain experts and are made publicly available on github repository[17]. A brief overview of different experiments is as given below:

## A. Baseline Approach

Our first experiment did not optimize indexing techniques for text collection. The document terms were unprocessed to generate index and subsequently perform retrieval and evaluation tasks. Detailed performance of each retrieval model is given in Pareek[18].

## B. Stop Word Removal

Stop words are the terms that frequently occur in a document but carry less significant meaning. In English language, words like *the, is, at, which, and,* etc. are considered as stop words. In most of the cases prepositions, conjunctions result in stop words. In our experiments, a set of 400 words were extracted that had highest frequency count in text collection and subsequently 282 words were considered as stop words with the help of linguistic experts who had manually inspected each word. In this experiment, we eliminated the stop words from index to perform IR tasks. A detailed discussion on experiments performed by eliminating stop words are discussed in Pareek[19].

## C. Stemming

Gujarati language is morphologically rich, which means that it involves many terms that can be of derived form or are inflected. Most of the cases, Gujarati terms are inflected with the use of suffixes. For better performance of IR systems, a technique called "stemming is applied", stemming is the process of reducing inflected terms into their root form. For instance, terms like "study", "studying", "studies", "studied", etc. can be reduced to the base form "study". In our experiments, we created a list of Gujarati suffixes for verbs, nouns, adjectives and adverbs and subsequently stemmed the terms to its root form for indexing of text documents. We experimented with statistical stemmer and rule based stemmer. Details of algorithm used for rule based stemmer can be found in Joshi, et al.[20]

## D. Frequent Case Generation (FCG)

Gujarati is a highly inflectional language[15] where one root can produce several morphological variants. Unlike English, proper nouns can also have a number of variations. In most of the cases, variants are generated by adding suffixes to the end of the root. There are six oblique forms in Gujarati that correspond to case forms nominative, genitive, instrumental, locative, accusative-dative. Gujarati verbs inflect for tense, aspect, mood, voice, person, number and gender. Therefore, Gujarati verbs agree with their subjects, adjectives inflect with gender, number, case and with nouns. However, adverbs do not inflect.

Kettunen and Airio[21] and Kettunen et al.[22] have developed a linguistic frequency based method call Frequent Case Generation (FCG) to generate variations of a given term. For instance, give a basic form "stem", the generator produces all the variant forms of it, in this case "stemmer", "stemmed", "stemming", "stems". These generated variants become input to the search engine which matches them in the plain inflected word form within the index.

## V. EVALUATION

Evaluation metrics are required to compare the performance of different IR systems. Although various metrics exist to evaluate IR systems, the widely used metrics are recall, precision and fallout[23]. Recall measures the fraction of relevant documents that are retrieved by the system whereas precision measures the fraction of retrieved documents that are relevant to the information need and fallout is a measure to indicate the fraction of non-relevant documents that are retrieved by the IR system. In the presence of multiple queries, a widely used metric, MAP (mean average precision)[24] is used to evaluate the performance of IR systems.

In our experiments, to evaluate the performance of IR systems, we used the evaluation metric, mean average precision (MAP) values for comparison of various IR systems. We evaluated the results separately for the queries tags with title (T), combination of title and description (TD) and the combination of title, description and narration (TDN). Mean average precision is derived from average precision (AP) values, where AP is the average of the precision values obtained for the topmost k number of documents retrieved. To calculate MAP, the AP values are then averaged over performance of multiple queries. For each query $q_j$ that belongs to a set of queries Q, we retrieve the set of relevant documents for that query denoted by a set $\{d_1, \ldots d_{mij}\}$ and let us assume that $R_{jk}$ is the set of ranked retrieval results topmost results until we get document $d_k$, then mean average precision (MAP) can be calculated as in (1).

$$\text{MAP (Q)} = \frac{1}{|Q|} \sum_{j=1}^{|Q|} \frac{1}{m_j} \sum_{k=1}^{m_j} \text{Precision (R}_{jk})  \quad (1)$$

## VI. RESULTS & ANALYSIS

We performed four categories of experiments with different indexing techniques using a set of 50 queries. Each experiment was performed on 20 different retrieval models to test the efficacy of these models on Gujarati language. The queries were varied by using combination of title (T), title & description (TD) and title, description & narration (TDN). Table 2 summarizes the results of each experiment.

TABLE 2
SUMMARY OF MAP VALUES FOR VARIOUS RETRIEVAL MODELS

| | T | TD | TDN |
|---|---|---|---|
| BaseLine | | | |
| Average | 0.217 | 0.232 | 0.152 |
| Max | 0.234 | 0.265 | 0.206 |
| Min | 0.194 | 0.126 | 0.018 |
| Stop words Elimination | | | |
| | T | TD | TDN |
| Average | 0.224 | 0.252 | 0.186 |
| Max | 0.237 | 0.269 | 0.220 |
| Min | 0.206 | 0.234 | 0.149 |
| Stemming | | | |

|         | T     | TD    | TDN   |
|---------|-------|-------|-------|
| Average | 0.223 | 0.259 | 0.212 |
| Max     | 0.233 | 0.275 | 0.238 |
| Min     | 0.202 | 0.243 | 0.190 |
| Frequent Case Generation (FCG) | | | |
|         | T     | TD    | TDN   |
| Average | 0.320 | 0.348 | 0.254 |
| Max     | 0.336 | 0.398 | 0.347 |
| Min     | 0.282 | 0.178 | 0.055 |

From the results as shown in Table 2, the following conclusions can be derived:

1) Combination of Title & Description (TD) in queries results in better precision rather than T or TDN.
2) It is observed that all the techniques have improved the MAP values of low performing models.
3) Applying stop word elimination, stemming and FCG techniques give better MAP values when compared to the baseline.
4) FCG technique improves the MAP values to nearly 67% of baseline for the TD case.

When we compare the performance of different retrieval models, it is observed that in most of the cases, the model In_expB2 (Inverse Expected Document Frequency model with Bernoulli after-effect and normalization 2)[14] is outperforming rest of the models resulting in highest MAP values for all experiments while the Hiemstra language model[10] falls below the baseline in most of the experiments.

## VII. CONCLUSION

On investigating the performance of each query within the set of 50 queries, it is observed that although bag-of-words approach is resulting in significant improvement of precision values, however, it does not take into account the semantics of query terms. For instance, a query "cat chases rat" and "rat chases cat" results in similar set of ranked retrieved documents. The bag-of-words approach is purely a statistical approach, additional techniques are required to restore the semantics with the text documents for efficient retrieval tasks.

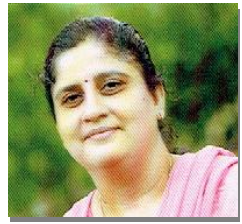

**Dr. Jyoti Pareek** is a Professor with Department of Computer Sc., Gujarat University She teaches students of MCA & M.Tech and is a PhD guide at Gujarat University & Banasthali University.She is a coordinator of MTech (NT) program at the department. She is a Senior IEEE member. Her research area includes Natural Language Processing, Software Engineering.

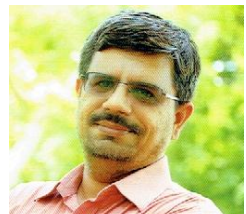

**Hardik Joshi** is Asst. Professor with the Department of Computer Sc., Gujarat University. He teaches students of MCA & M.Tech of Gujarat University. He is a coordinator of M.Tech(WT) program and Animation department of Gujarat University. He is a member of ACM and his research area includes Natural Language Processing & Information Retrieval.

**Krunal Chauhan** is a Sr. Software Engineer at LogiCeil Solutions, Ahmedabad, India and **Rushikesh Patel** is a Solution Analyst at Canada Technology Partners Ltd., India. Both Krunal & Rushikesh work in the area of Big Data Analytics on Industrial Projects.